\documentstyle[12pt]{article}
     \def\lsim{\raise0.3ex\hbox{$<$\kern-0.75em\raise-1.1ex\hbox{$\sim$}}}
\def\gsim{\raise0.3ex\hbox{$>$\kern-0.75em\raise-1.1ex\hbox{$\sim$}}}
\def\noi{\noindent}
\def\nn{\nonumber}
\def\bea{\begin{eqnarray}}  \def\eea{\end{eqnarray}}
\def\beq{\begin{equation}}   \def\eeq{\end{equation}}

\begin{document}
\begin{center}
\vbox to 1 truecm {}
{\Large \bf Absence of anomalous stopping in heavy ion collisions} \\

\vskip 8 truemm
{\bf A. Capella}\\
\vskip 5 truemm

{\it Laboratoire de Physique Th\'eorique\footnote{Unit\'e Mixte de
Recherche UMR n$^{\circ}$ 8627 - CNRS}
\\ Universit\'e de Paris XI, B\^atiment 210,
F-91405 Orsay Cedex, France} \\
\end{center}

\vskip 1 truecm

\begin{abstract}
We show that the baryon stopping observed in heavy ion collisions
both at CERN-SPS and at RHIC can be derived from the one observed in
proton-proton
collisions. No increase in the size of the baryon junction component
is required between small size $(pp)$ and large size $(AA)$ systems.
\end{abstract}

\vskip 3 truecm

\noi LPT Orsay 02-43 \par
\noi May 2002\par
\newpage
\pagestyle{plain}
A huge baryon stopping has been observed in central $Pb$ $Pb$
collisions at CERN-SPS by the NA49 collaboration \cite{1r}. This
stopping cannot be
obtained from the standard fragmentation of a diquark. In view of
that, several authors [2-4] have introduced a mechanism related to
the transfer in
rapidity of the string junction (SJ), which carries the baryon
number. This mechanism had been considered long time before the CERN
heavy ion program
started \cite{5r,6r}. In ref. \cite{6r} it was argued that such a
mechanism was required in proton-proton collisions in order to
explain the net proton
production at mid-rapidities. This has been confirmed recently in
ref. \cite{7r}. In particular, the NA49 data \cite{8r} on the
inclusive reaction $\pi +
p \to (p - \overline{p}) + X$ in the pion hemisphere, analysed in
\cite{7r}, show very clearly the necessity of the SJ mechanism. This
mechanism will be
described in detail in the next section. \par

In most string models of multiparticle production, in particular the
Dual Parton Model (DPM) \cite {9r} and the Quark Gluon String Model
\cite{10r},
the net baryon production at mid-rapidities increases with the number
of inelastic collisions suffered by the nucleon from which it
originates. Indeed,
in the case of production via diquark fragmentation, when the number
of inelastic collisions increases the diquark gets slower due to
energy momentum conservation. Hence, the produced net baryon is also
slower. The same increase of stopping with the number of inelastic
collisions occurs in
the case of production via the SJ mechanism. It is then natural to
ask whether the stopping observed in heavy ion collisions, both at
CERN-SPS
\cite{1r} and at RHIC \cite{11r,12r} is normal or anomalous. More
precisely, from hadron-hadron scattering data we can determine the
strength of the
component associated to the SJ mechanism. Using this
value, one can then compute the
net baryon production in heavy ion collisions, where the average
number of inelastic
collisions per participant is larger. If the data can be reproduced
in this way, the observed stopping can be considered normal. If, on
the contrary, the
heavy ion data can only be explained by increasing the coupling of
the SJ component, the observed stopping can be considered anomalous.
\par

In this paper we show that present data can be explained with the
same strength of the SJ component in $pp$ and central $AA$ collisions
and, thus, they give
no indication of anomalous stopping. \\

\noi \underbar{\bf SJ transfer mechanism}. In string models there are
three different mechanisms of net baryon production. In the
first one, production takes place from the fragmentation of a valence
diquark. In this case the baryon junction follows the diquark, which
fragments into a leading baryon by picking up a sea quark out of the
$q$-$\overline{q}$ pairs produced in the string breaking. The
produced net baryon is
thus made out of two valence and one sea quarks. This mechanism will
be denoted diquark preserving (DP). \par

In the other two cases the diquark is broken (DB). In one case the
string junction travels in rapidity together with a valence quark.
The produced net
baryon is made out of one valence and two sea quarks. In the other
case, the string junction travels in rapidity without any valence
quark (gluonic
mechanism) and the produced net baryon is made out of three sea quarks.\par

The important point is to determine the rapidity distribution
associated to each of these three mechanisms. In a Regge language
this amounts to
determining the Regge intercept associated to the exchanged objects~:
$SJ$, $SJ+q_v$ and $SJ+(qq)_v$. Two different choices for these
intercepts
have been considered in the literature for the DB mechanisms~:

\beq
\label{1e}
\alpha_{SJ}(0) = 1/2 \quad , \quad \alpha_{SJ+q_v}(0) = 0\
\cite{5r}\footnote{One of the authors of ref. \cite{5r} (G.V.)
considers presently the choice
(2) as very natural (G. Veneziano, private communication).} \eeq

\noi and

\beq
\label{2e}
\alpha_{SJ}(0) = 1\quad , \quad \alpha_{SJ+q_v}(0) = 1/2 \ \cite{6r} \ .
\eeq

\noi Note the equal spacing in eqs. (\ref{1e}) and (\ref{2e}) due to
standard Regge rules~: $\alpha_{SJ+q_v}(0) = \alpha_{SJ}(0) +
\alpha_{\rho}(0) - 1$,
where $\alpha_{\rho}(0) = 1/2$.\par

Experimentally, there is ample evidence for the existence of a
component with intercept 1/2 in hadron-hadron scattering \cite{7r}.
In these data
there is no room for a component with intercept 1 \footnote{With the
possible exception \cite{13r} of preliminary HERA data \cite{14r}.
See also ref.
\cite{7r} for a discussion on this point.}. Net baryon production in
heavy ion collisions, both at CERN-SPS and RHIC, also requires a DB
mechanism
with intercept 1/2 [2-4, 15, 16]. Moreover, the values of the ratios
$\overline{B}/B$ measured at RHIC indicate that the corresponding net
baryon is
(dominantly) made out of one valence and two sea quarks \cite{16r}
\footnote{Note, however, that a non-zero value of net omegas has been
observed in $hA$ collisions \cite{17r}. This requires a non-vanishing
contribution in which the net baryon is made out of three sea quarks.
However, its effect in $AA$ collisions is probably very small since,
in this case,
the produced net omegas are almost entirely due to final state
interaction \cite{15r} \cite{16r}.}. This clearly favors possibility
(\ref{2e}), which will be used throughout this paper.
This probably indicates that the gluonic SJ component has a coupling
too small to manifest itself at present energies, where the
pre-asymptotic component
$SJ + q_v$ with intercept 1/2 dominates \cite{3r}.\par

Let us turn next to the DP mechanism $SJ + (qq)_v$, i.e. the
conventional diquark (plus SJ) fragmentation. Its intercept
corresponds to the so-called
baryonium exchange, whose intercept is known experimentally to be
$-1.5 \pm 0.5$. We will use this experimental value in what follows.
Note
that, according to the usual Regge rules, this intercept should have
equal spacing with respect to the ones in (\ref{1e}) and (\ref{2e}),
i.e.
should have intercept $- 1/2$ in case (\ref{1e}) and 0 in case
(\ref{2e}). The same Regge rules allow to relate these intercepts to
the intercept of
the nucleon trajectory, $\alpha_N(0)$ \cite{6r} \cite{18r}
\cite{19r}. Thus, $\alpha_{SJ+(qq)_v}(0) = - 1/2$, (0) corresponds to
$\alpha_N(0) = 0$ (1/4),
whereas the experimental value $\alpha_{SJ + (qq)_v}(0) = - 1.5 \pm
0.5$ correspond to $\alpha_N(0) = - {1 \over 2} \pm {1 \over 4}$.
These discrepancies\footnote{Note also that in QGSM a value
$\alpha_{SJ+q_v}(0) = - 1$ is used \cite{18r} \cite{19r}, obtained
from the experimental value $\alpha_{SJ + (qq)_v}(0) = - 1.5$, using
the 1/2 Regge spacing rule
mentioned above. In this case a component with intercept 1/2
corresponds to the gluonic SJ exchange
and the 1/2 -- equal --
spacing rule is broken between $\alpha_{SJ}(0) = 1/2$ and $\alpha_{SJ
+ q_v}(0) = - 1$.}
are probably related to spin effects responsible for the different
intercepts of $N$ and $\Delta$ trajectories. \\

\noi \underbar{\bf The model}.  Our model for net baryon production
consists of two different components associated to the two mechanisms
described above.
A conventional (DP) component corresponding to the fragmentation of a
valence diquark and a DB component in which the SJ follows one of the
two valence
quarks of the broken diquark. As discussed above, in the first case
the baryon is made out of two valence and one sea quarks and is
mostly produced in
the fragmentation region. In the second case, it is made out of two
sea and one valence quarks. This second component gives the dominant
contribution at
mid-rapidities.
In DPM, QGSM and most string
models the hadron spectra
of the individual strings are obtained as convolutions of momentum
distribution and fragmentation functions. However, in the case of net
baryon production
it is natural to assume that its rapidity distribution is essentially
the one of the SJ, which is given by its Regge intercept. In this way
we are led to
the following model\footnote{Note that eq. (\ref{3e}) ensures exact
baryon number conservation. This conservation is not so easy to
enforce when a convolution of momentum
distribution and fragmentation functions is performed. However, eq.
(\ref{3e}) does not keep track of the fact that the string in which
the net baryon is produced has a
valence quark at the opposite end. The average value of the rapidity
of the latter is about 1.5 units from the maximal value of the
rapidity of the proton to which it
belongs. Consequently, eq. (\ref{1e}) can not be used to compute the
net baryon contribution at the end of phase space in the opposite
hemisphere. However, this is of no
consequence for the results on the total net baryon distribution in
$pp$ or $AA$ collisions since the contribution to the fragmentation
region of a nucleon from the
nucleon on the opposite side is very small as compared to the one
from the same side.} for the net baryon production out of a single
baryon

\bea
\label{3e}
&&{dN_{\mu (b)} \over dy}(y) = a C_{\nu}^{DB} Z_+^{1 -
\alpha_{SJ+q_v}(0)} (1 - Z_+)^{\mu (b) - 3/2 + n_{sq} 
(\alpha_{\rho}(0) - \alpha_{\phi}(0))} \nn \\
&&+ (1 - a)
C_{\nu}^{DP} Z_+^{1 - \alpha_{SJ + (qq)_v}(0)} (1 - Z_+)^{\mu (b)  -
3/2 + c + n_{sq} (\alpha_{\rho}(0) - \alpha_{\phi}(0))} \eea

\noi where $n_{sq}$ is the number of strange quarks in the hyperon
$\alpha_{\rho}(0) = 1/2$ $\alpha_{\phi}(0) = 0$, $Z_+ =
(e^{y-y_{max}})$, $y_{max}$ is the
maximal value of the baryon rapidity and $\mu (b) $ is the average
number of inelastic collisions suffered by the baryon at fixed impact
parameter $b$ \footnote{Actually
the number of inelastic collisions is distributed. However, the
distribution is rather narrow and no significant numerical difference
has been observed by taking the
average value, as in eqs. (\ref{3e}).}. The constants $C_{\nu}$ are
obtained from the normalization to unity of each term. The small $Z$
behaviour is controled by
the corresponding intercept. The factor $(1 - Z_+)^{\mu (b) - 3/2}$
is obtained by requiring that the $Z$-fractions of all quarks at the
ends of the strings,
other than the one in which the baryon is produced, go to zero
\cite{15r}. Following conventional Regge rules \cite{18r,19r,7r} an
extra $\alpha_{\rho}(0) =
\alpha_{\phi}(0) = 1/2$ is added to the power of $1 - Z_{+}$ for each
strange quark in the hyperon. \par

The fraction, $a$, of the DB breaking component is treated as a free
parameter. The same for the parameter $c$ in the DP component --
which has to be
determined from the shape of the (non-diffractive) proton inclusive
cross-section in the baryon fragmentation region. In the case of $AA$
collisions the
value of $\mu (b) $ is given by $\mu (b) = k \nu (b)$ with $\nu (b) =
n(b)/n_A(b)$, where $n(b)$ and $n_A(b)$ are the average number of
binary collisions and of participant
of nucleus $A$ and $k$ is the average number of inelastic collisions
in $pp$. At SPS energies $k = 1.4$ and at RHIC $k = 2$ \cite{20r}.
\par

Related models have been proposed in \cite{3r} and \cite{15r}. The
results for heavy ion collisions are rather similar to the ones
obtained from Eq. (3).
However, in these models there is some increase in the size of the DB
component with the number of inelastic collisions -- suggesting an
anomalous
stopping. In a very recent paper \cite{sha} it is found that anomalous
stopping is also needed in the model of ref. \cite{7r}.\\

\noi \underbar{\bf Quark counting rules}. Eqs. (\ref{3e}) do not
allow to determine the relative densities of the different baryon
species. In order to
do so we use simple quark counting rules \cite{15r}, \cite{16r}. Let
us denote the strangeness suppression factor by $S/L$, with $2L + S =
1$. Baryons
produced out of one valence and two sea quarks, which is the case for
the DB component, are given the relative weights $I_2 = 2L^2 :
2L^2:4LS:S^2/2:S^2/2$
for $p$, $n$, $\Lambda + \Sigma$, $\Xi^0$ and $\Xi^-$, respectively.
The various coefficients of $I_2$ are obtained from the power
expansion of
$(2L+S)^2$. For baryons made out of two valence and one sea quark,
which is the case for the DP component, the corresponding weights are
given by $I_1 =
L:L:S$ for $p$, $n$ and $\Lambda + \Sigma$, respectively. In the
calculation we use $S = 0.1$ ($S/L = 0.22$) as in ref. \cite{16r}.
The above weights
apply to the case where the number of valence $u$ and $d$ quarks is
the same. In the case of an incoming proton they have to be modified
in a
straighforward way. For example, in the DB (DP) component, the weight
for the production of a net proton is changed from
$2L^2(L)$ into
$7L^2/3$ $(4L/3)$. Likewise the modification due to the fact that
about 60 \% of the nucleons of the nucleus are neutrons is
straightforward. \\

\noi \underbar{\bf Numerical results}. A good description of the data
on the rapidity distribution of $pp \to p-\overline{p} + X$ both at
$\sqrt{s} = 17.2$~GeV \cite{21r}
and $\sqrt{s} = 27.4$~GeV \cite{22r} is obtained from eq. (\ref{3e})
with $a = 0.4$, $c = 1$, $\alpha_{SJ+q_v} = 1/2$ and
$\alpha_{SJ+(qq)_v} = - 1$. The results are
shown in Table 1 at three different energies, and compared with
the data \cite{21r,22r}. As we see the agreement is reasonable.
We have also checked that eq.
(\ref{3e}) reproduces the preliminary data of the NA49 collaboration
on $pp \to p-\overline{p} + X$ \cite{23r} and $\pi p \to
p-\overline{p} + X$ \cite{8r}, as well as the
ones on $pp \to \Lambda - \overline{\Lambda} + X$ \cite{19r}. For
comparison with the nucleus-nucleus results, all values in Table 1
have been scaled by the
number of participants pairs in central $Pb$ $Pb$ and $Au$ $Au$
collisions ($n_A = 175$). As it is well known, a pronounced minimum
is present at $y^* = 0$. There is
also a substantial decrease of the mid-rapidity yields with
increasing energy. Also, the mid-rapidity distributions get flatter
with increasing energy
since the net proton peaks are shifted towards the fragmentation
regions.\par

It is now possible to compute the corresponding net baryon production
in heavy ion
collisions and to check whether or not the data can be described with
eq. (\ref{3e}) using the same set of parameters.  \par

The results for net proton ($p - \overline{p}$) and net baryon ($B -
\overline{B}$) production in central $Pb$ $Pb$ collisions at
$\sqrt{s} = 17.2$~GeV and central $Au$ $Au$ collisions at $\sqrt{s} =
130$~GeV are given in Table 2. The centrality is defined by the
average number of participants --
$n_{part} = 350$ in both cases. Experimental results \cite{1r}
\cite{11r} are given in brackets.\par

The comparison of column 2 with the $pp$ results in 
Table 1 at the same energy,  shows the
well known change in the shape of the rapidity distribution between
$pp$ and central $Pb$ $Pb$
collisions. The minimum at $y^* = 0$ is much less pronounced in $Pb$
$Pb$ and the net proton peaks in the $pp$ fragmentation regions are
shifted to $y^* \sim \pm 1.5$.
More interesting are the results in columns 4 and 5 which contain the
predictions at RHIC (where data are only available at $y^* \sim 0$).
We see that the 
shape of the rapidity distribution is very different from the one
at SPS. (At RHIC the value at $y^* = 0$ is
smaller than at SPS by a factor 3, whereas at $y^* = 2$ the decrease
is only 45 \%). We also see
that the peaks at $|y^*| = 1.5$ are not present at RHIC, (i.e. they
are shifted to higher values of $|y^*|$). \par

The main result of our work is the fact that the calculated values in
$Pb$ $Pb$ and $Au$ $Au$ collisions are in reasonable agreement with
experiment. Actually, the
calculated value at RHIC is even higher than the experimental
value\footnote{It has been shown in \cite{15r} \cite{16r} that final
state interaction is crucial in
order to obtain the observed enhancement of strange baryons both at
SPS and RHIC. Due to strangeness conservation, this produces a small
decrease of the net proton yield
-- while the net baryon yield is not changed. This decrease is
maximal at $y^* = 0$, where it is of 14 \% at RHIC \cite{16r}. The
computed value for $Au\, Au \to p -
\overline{p}$ is thus reduced from 8.0 to 7.0 -- in better agreement
with experiment.}. This has been achieved using the same parameters
as in $pp$ collisions. In
particular, there is no need to increase the value of the parameter
$a$, which determines the strength of the SJ component. \par

In conclusion, we have introduced a two component model for net
baryon production in $pp$, $pA$ and $AA$ collisions. One component
corresponds to the
conventional diquark fragmentation mechanism and produces baryons
mostly in the fragmentation regions. The other component is
associated with the mechanism
of transfer in rapidity of the baryon junction and gives the dominant
contribution at mid-rapidities. The model allows to compute the net
baryon rapidity
distribution for the different baryon species and for all
centralities at any given energy. The model contains two free
parameters -- the most important
one being the parameter $a$ which determines the relative strength
(40 \%) of the baryon junction component. We have shown that the same
set of parameters
allows to describe the data on net proton production in $pp$ and $AA$
collisions. This leads to the conclusion that no anomalous stopping
has been
observed in heavy ion collisions.\\

\begin{center}
\begin{tabular}{|c|c|c|c|}
\hline
$y^*$ &$pp \to p - \overline{p}$ &$pp \to p - \overline{p}$ &$pp \to
p - \overline{p}$ \\
&$\sqrt{s}= 17.2$ GeV &$\sqrt{s} = 27.4$ GeV &$\sqrt{s} = 130$ Gev\\
\hline
0 &9.2 &6.5 &3.6 \\
  & &$[6.3 \pm 0.9]$ & \\
1 &15.0 &9.3 &4.2 \\
  &$[16.1\pm 1.8]$ &$[9.6 \pm 0.9]$ & \\
1.5 &25.8 &14.6 &5.1\\
&$[24.1 \pm 1.4]$ &$[15.4 \pm 0.9]$ & \\
2 &47.1 &26.2 &6.8 \\
&$[45.4 \pm 1.4]$ &$[27.7 \pm 0.9]$ & \\
\hline
\end{tabular}
\end{center}
\vskip 5 truemm

\noi {\bf Table 1.} Calculated values of the rapidity distribution of
$pp \to p - \overline{p} + X$ at $\sqrt{s} = 17.2$~GeV and 27.4 GeV
($k = 1.4$) and $\sqrt{s} =
130$~GeV ($k=2$). The data in the second column are from ref.
\cite{21r}. (In order to convert $d\sigma/dy$ into $dN/dy$ a value of
$\sigma = 30$~mb has been used). The
data at $\sqrt{s} = 27.4$ GeV are from ref. \cite{22r}, as presented
in \cite{1r}. Following \cite{1r}, and for comparison with the
nucleus-nucleus results, all values in
Table 1 have been scaled by $n_A = 175$ -- the number of participant
pairs in central $Pb$ $Pb$ and $Au$ $Au$ collisions. \\

\begin{center}
\begin{tabular}{|c|c|c|c|c|}
\hline
$y^*$ &$Pb\, Pb \to p - \overline{p}$ &$Pb\, Pb \to B - \overline{B}$
&$Au\, Au \to p - \overline{p}$ &$Au\, Au \to B - \overline{B}$\\
&$\sqrt{s} = 17.2$ GeV &$\sqrt{s} = 17.2$ GeV &$\sqrt{s} = 130$ GeV
&$\sqrt{s} = 130$ GeV\\
\hline
  0 &23.0 &58.5 &8.0 &20.9 \\
&$[26.7 \pm 3.7]$ &$[67.7 \pm 7.3]$ &$[5.6 \pm 0.9 \pm 24 \%]$ & \\
1 &32.3 &79.7 &8.6 &22.6 \\
&$[34.9 \pm 1.5]$ &$[84.7 \pm 3.5]$ & & \\
1.5 &36.3 &87.0 &12.3 &31.5 \\
&$[34.4 \pm 1.7]$ &$[80.0 \pm 3.9]$ & & \\
2 &25.3 &57.15 &17.3 &43.4 \\
&$[24.7 \pm 1.5]$ &$[56.1 \pm 3.1]$ & & \\
\hline
\end{tabular}
\end{center}
\vskip 5 truemm

\noi {\bf Table 2.} Calculated values of the rapidity distribution
$dN/dy$ for central $Pb$ $Pb \to p - \overline{p} + X$ and $Pb$ $Pb
\to B - \overline{B} + X$ at
$\sqrt{s} = 17.2$~GeV ($k = 1.4$) and central $Au$ $Au \to p - \overline{p} + X$
and $Au$ $Au \to B - \overline{B} + X$ at 
$\sqrt{s}= 130$~GeV ($k = 2$). The
centrality has been defined by the
number of participant pairs ($n_A = 175$ at both energies) and $\nu =
n/n_A = 4.5$ (5.0) at SPS (RHIC). Data are from refs. \cite{1r} and
\cite{11r}.

\vskip 1.5cm

\noi \underbar{\bf Acknowledgments}. It is a pleasure to thank
G. Arakelyan, H. G. Fischer, A. Kaidalov, B. Kopeliovich,
C. Salgado, Yu. Shabelski, D. Sousa and G. Veneziano for discussions.
This work was supported in part by the European Community-Access to
Research Infrastructure action of the Improving Human
Potential Program.

\end{document}